\documentclass[journal]{IEEEtran}
\usepackage{amsmath,amsfonts}
\usepackage{algorithmic}
\usepackage{algorithm}
\usepackage{array}
\usepackage[caption=false,font=normalsize,labelfont=sf,textfont=sf]{subfig}
\usepackage{textcomp}
\usepackage{stfloats}
\usepackage{url}
\usepackage{verbatim}
\usepackage{graphicx}
\usepackage{cite}
\begin{document}

\title{Federated Deep Reinforcement Learning-Based Intelligent Channel Access in Dense Wi-Fi Deployments}

\author{
Xinyang Du, Xuming Fang,~\IEEEmembership{Senior Member,~IEEE,} and Rong He, Li Yan, Liuming Lu, Chaoming Luo
\thanks{The work of X. Du, and X. Fang was supported in part by the NSFC under Grant No. 62071393. (Corresponding author: Xuming Fang).}
\thanks{X. Du, X. Fang, R. He and L. Yan are with Key Lab of Information Coding \& Transmission, Southwest Jiaotong University, Chengdu 610031, China (e-mails: xydu@my.swjtu.edu.cn; xmfang@swjtu.edu.cn; rhe@home.swjtu.edu.cn; liyan@swjtu.edu.cn).

L. Lu and C. Luo are with Guangdong OPPO Mobile Telecommunications Corp., Ltd, GuangDong 523859, China (e-mails: luliuming@oppo.com; luochaoming@oppo.com).
}
}



\maketitle

\begin{abstract}
The IEEE 802.11 MAC layer utilizes the Carrier Sense Multiple Access with Collision Avoidance (CSMA/CA) mechanism for channel contention, but dense Wi-Fi deployments often cause high collision rates.  To address this, this paper proposes an intelligent channel contention access mechanism that combines Federated Learning (FL) and Deep Deterministic Policy Gradient (DDPG) algorithms. We introduce a training pruning strategy and a weight aggregation algorithm to enhance model efficiency and reduce MAC delay. Using the NS3-AI framework, simulations show our method reduces average MAC delay by 25.24\% in static scenarios and outperforms A-FRL and DRL by 25.72\% and 45.9\% in dynamic environments, respectively.
\end{abstract}


\section{Introduction}
\IEEEPARstart{W}{ireless} Local Area Network (WLAN) technology has been widely used in various scenarios such as smart home, Internet of things and so on, due to its characteristics of high data rate, low latency, easy deployment and configuration. To meet the growing demand for high-quality data transmission, the IEEE established the 802.11bn (i.e. Wi-Fi 8) Task Group in July 2022 \cite{b1}, aiming to develop ultra-reliable wireless networks. However, in dense environments where a large number of devices are accessing the network, the likelihood of collision increases during multi-user channel contention, leading to data loss or transmission failures, which adversely affect throughput and latency performance.

The existing 802.11 standard employs the Distributed Coordination Function (DCF) or Enhanced Distributed Channel Access (EDCA) mechanism, which combines Carrier Sense Multiple Access with Collision Avoidance (CSMA/CA) and the Binary Exponential Backoff (BEB) mechanism for channel contention \cite{b2}. The EDCA mechanism improves overall network performance by providing differentiated Quality of Service (QoS) for various types of traffic. This is achieved by prioritizing high-priority traffic through the configuration of different contention window sizes and Arbitration Inter-Frame Space (AIFS) parameters, and it is widely used in Wi-Fi scenarios.

While traditional methods of adjusting the contention window(CW) can reduce collisions to some extent, they may not be optimal in all cases, potentially leading to unnecessary backoff times and inefficient use of channel resources. Bianchi et al. \cite{b3} analyzed the performance of CSMA/CA using a two-dimensional Markov chain and introduced the DCF Request To Send/Clear To Send (RTS/CTS) mechanism to improve system performance to a certain extent. Zhang et al. \cite{b4} proposed an adaptive adjustment of the contention window update factor to address the issue where a fixed update factor fails to adapt to changes in network scale. Rehman et al. \cite{b5} introduced a collision-based window scaling backoff mechanism, which enhances channel resource allocation efficiency by selecting an optimal window size for each collided or successfully transmitted data frame.

However, despite the improvements in system throughput achieved by optimizing the BEB mechanism, the increasing density and complexity of WLAN networks have rendered the performance of traditional optimization algorithms unstable, failing to meet the Quality of Experience (QoE) demands of users. Therefore, it is crucial to explore an efficient, practical, and robust solution tailored to the 802.11 MAC layer mechanisms.

In the research focused on optimizing IEEE 802.11 MAC layer mechanisms to enhance Wi-Fi network performance, machine learning methods have been introduced. Specifically, most studies leverage Q-learning and multi-armed bandit algorithms for optimizing channel access, with the aim of improving throughput and QoS \cite{b6,b7,b8,b9}. Deep Reinforcement Learning (DRL), as a powerful machine learning technique, has been extensively applied for the dynamic adjustment of MAC layer parameters such as the contention window, owing to its efficient decision-making capabilities in complex environments \cite{b10,b11,b12,b13}. Meanwhile, the architectures used in these studies are primarily categorized into single-agent and distributed multi-agent. Single-agent architectures are suitable for simple scenarios \cite{b11,b14}, whereas distributed multi-agent architectures exhibit better scalability and robustness in densely deployed environments \cite{b10,b15}. These studies typically focus on performance metrics such as network throughput, latency, and fairness, with fairness as an increasingly important criterion for evaluating algorithmic performance, especially in multi-user environments. Nevertheless, the centralized training methods involved in these studies often incur substantial over-the-air overhead and raise concerns regarding the data privacy of STAs.

Given that wireless network environments are typically characterized by distributed and dynamic properties, relying solely on the local learning models of STAs may not adequately capture the global information of the entire network. This limitation can lead to a degradation in overall network performance and reduced fairness among devices. To address these challenges, McMahan et al.  \cite{b16} proposed the application of Federated Learning (FL) for optimizing communication networks. FL allows individual network STAs to train models locally and upload the model parameters to a central server for aggregation. This approach provides more accurate and reliable model training and estimation in dynamic and frequently changing wireless network environments, reduces communication overhead, and enhances the robustness and accuracy of the model. Consequently, FL has emerged as an effective solution to the issues faced by centralized training approaches in current wireless networks. Zhuang et al. \cite{b17} proposed a novel enhanced multi-access mechanism based on Deep Reinforcement Learning (DRL) and FL, achieving fairness among users through the application of FL. Ali et al. \cite{b18,b19} introduced a collaborative machine learning algorithm utilizing the Federated Reinforcement Learning (FRL) framework, where the decentralized strategy based on FRL improved system fairness and reliability.

However, in the aforementioned schemes, FL typically adopts an averaging algorithm for aggregation, with all STAs participating in the process, without considering the variations in STA-to-AP distance, channel quality, and sample quantity that may arise in dynamically dense distributed Wi-Fi environments. These differences can result in certain STAs being unsuitable for FL model training. If STAs with insufficient training time or poor-quality samples participate in the aggregation, the overall model performance may be compromised. To enhance the FL efficiency and ensure data diversity and model generalization capability, it is necessary to design a reasonable pruning strategy for FL model training.

To address these challenges, we propose an Exponential Aggregation-based Federated Reinforcement Learning (E-FRL) intelligent channel contention access method, along with a pruning strategy to alleviate the burden on global aggregation during model training. The specific contributions are as follows:
\begin{itemize}
\item We propose a distributed algorithm based on a single-AP multi-STA dense Wi-Fi scenario, utilizing an FL architecture as the core. In this algorithm, each agent deploys a Deep Deterministic Policy Gradient (DDPG) model for intelligent decision-making, thereby enhancing the fairness and stability of system access.
\item In response to the dynamically changing scenarios of channel quality and STA uplink traffic, we design a training pruning strategy based on state information such as STA distance and packet loss rate. We also propose an exponential aggregation algorithm to address issues caused by poor training samples, improving the training efficiency and accuracy of the model, and ensuring the robustness and adaptability of the model under varying network conditions.
\end{itemize}

\section{System Model}
 Assume a Wi-Fi network consisting of an Access Point (AP) and $N$ associated Stations (STAs), denoted as $\mathcal{N}=\{1,2,3,\cdots,N\}$. Time is divided into equal-length slot periods, with the interaction time step denoted by \(t\). For the sake of analysis, this paper considers the uplink transmission as an example, using Single-User Multiple Input Multiple Output (SU-MIMO) for data transmission. It is assumed that each STA deploys an agent capable of independent model training and sample storage, while the AP functions as the server for FL, responsible for initiating and coordinating the entire FL process.


According to \cite{b20}, the variation in contention windows in a Wi-Fi scenario follows a Markov Decision Process (MDP). Therefore, this paper models the intelligent adjustment of contention windows based on the MDP framework. The DDPG algorithm is an ideal choice for addressing Wi-Fi contention window optimization problems due to its ability to handle continuous action spaces and its adaptability to complex environments.

The key to solving RL problems lies in constructing the agent and environment, which involves designing parameters such as state, action, and reward. Considering that the size of the contention window is closely related to the channel state and user density, we define the packet loss rate (PLR) and the idle channel time ratio as the states of the RL model, denoted as \( \quad s_t = \{PLR_t, Idle_t\}\).The frame transmission failure probability \({PLR}_t\) is calculated by the STA according to \eqref{e_1}, considering only uplink transmission, where \(N_{tx}\) is the number of frames sent by the STA within an interaction period \(t\), and \(N_{ack}\) is the number of ACKs received by the STA.
\begin{equation}
\label{e_1}
    PLR_t = \frac{N_{tx} - N_{ack}}{N_{tx}}
\end{equation}

The idle channel time ratio \(Idle_t\) is a critical indicator for evaluating the performance of wireless communication systems. When the traffic load is fixed, it can reflect the intensity of channel contention to a certain extent. Let \(T_{OBS}\) denote the observation interval and \(T_{Busy}\) denote the time the channel is busy, then \(Idle_t\) can be expressed as:
\begin{equation}
\label{e_2}
    Idle_t = \frac{T_{OBS} - T_{Busy}}{T_{OBS}}
\end{equation}

In this paper, we optimize the average MAC delay performance of system by adjusting the upper limit of the contention window range. The STA randomly selects a backoff window value within the range \([0,CW_t]\), where the relationship between \(CW_t\) and \(a_t\) is given by:
\begin{equation}
\label{e_3}
    CW_t = \left\lfloor 2^{a_t + 4} - 1 \right\rfloor
\end{equation}

Given that the protocol defines the CW range as 15 to 1023, the action space \(a_t\) can be represented as \(a_{t}=\{0,1,2,3,4,5,6\}\). Since the DDPG algorithm operates in a continuous action space, the model's output needs to be discretized. Specifically, the continuous output value of DDPG is multiplied by 6 and rounded to map the continuous action to a discrete contention window value.

The design of the reward function primarily considers the delay performance. For ease of analysis, the delay is normalized by introducing a delay threshold \(delay_{limit}\) and set \(delay_{limit}\) be 20ms. Let \(delay_t\) denote the average MAC transmission delay of all data packets within an interaction time \(t\), then the reward function can be expressed as:
\begin{equation}
\label{e_4}
    r_t = \frac{delay_{limit} - delay_t}{delay_{limit}}
\end{equation}

\section{Adaptive Contention Window Adjustment Scheme Based on the E-FRL Algorithm}
\subsection{FRL Framework }
The E-FDL algorithm adopts an FL framework, where the FL process is periodic, primarily consisting of a local training phase and an aggregation phase. During the local training phase, each STA interacts with its environment independently, completing sample storage and model training. In the aggregation phase, the AP selects STAs for aggregation based on state parameters such as distance and channel quality. The selected STAs upload their model weight parameters to the AP, which then calculates the model parameters using a weighted aggregation algorithm defined by itself, resulting in a global model. This global model is subsequently distributed to the STAs participating in that round of aggregation. It is important to note that, given the relatively small data size of the model parameters, the overhead of uploading these parameters to the AP and the subsequent distribution of the global model to each STA is neglected in this paper. 

By introducing the FL framework, we aggregate the parameters of each agent within the distributed algorithm, thereby preventing unfair competition among STAs that could result in overly aggressive decision-making. The inclusion of FL not only enhances the overall fairness of the system but also optimizes resource allocation among STAs, thereby improving the system’s latency performance as a whole.

\subsection{Model Pruning Strategy for Training}
In consideration of the quality of the training model, the system employs a model pruning strategy during training to enhance the efficiency and accuracy of model training. The purpose is to optimize STA selection during training to achieve a higher quality model.

We primarily consider two factors in selecting STAs: the distance between the STA and the AP, and the historical average packet loss rate. The steps in the selection mechanism include:

\begin{itemize}
\item Sample Quantity Filtering: If a STA's sample size is less than the minimum required for training, that STA will not participate in model aggregation. This ensures that STAs participating in aggregation have a sufficient number of samples, thereby maintaining the quality of the training data.
\item Comprehensive Scoring Mechanism: To optimize the quality of model training, we define a comprehensive scoring mechanism to select the optimal STAs. The scoring takes into account the distance between the STA and the AP, as well as the STA’s historical packet loss rate. The number of STAs participating in aggregation can be adjusted according to the requirements of different scenarios.
\end{itemize}

The distance \(d_n\) between the STA and the AP affects signal strength and transmission quality. The farther the distance, the more severe the signal attenuation. Therefore, the score for distant STAs should be reduced, which is reflected in the distance score\(\quad S_n^d = \frac{d_{max}-d_n}{d_{max}}\), where \(d_{max}\) is the maximum coverage distance of the AP. The historical packet loss rate \({PLR}_n\), defined as the average packet loss rate over the past ten slots, can generally reflect the stability of the STA’s communication. The higher the packet loss rate, the lower the score should be. Therefore, the packet loss rate score is defined as \(S_n^{PLR}=1-\ {PLR}_n\). The comprehensive score can be expressed as:
\begin{equation}
    S_n = w_1 S_n^d + w_2 S_n^{PLR}
\end{equation}

where \(w_1\) and \(w_2\) are weight parameters used to balance the impact of distance and packet loss rate, with the condition that \(w_1 + w_2 = 1\).This Pruning Strategy helps in choosing STAs with higher signal quality and stability, thereby improving the efficiency and accuracy of model training.

\subsection{Exponential-Based Aggregation Method}
In most FL schemes aimed at Wi-Fi optimization, a simple averaging method is typically used to aggregate local models to form a global model. However, this kind of approach does not fully consider the impact of the number of samples from each STA on the aggregation process, which may lead to a decline in model performance. In supervised learning, the FedAvg aggregation mechanism is commonly used \cite{b16}, where the aggregation model is weighted based on the proportion of each STA's samples to the total sample size. However, this direct weighting method based on sample size also has limitations and may not fully address the issue of aggregation interference caused by insufficient sample sizes. This is especially problematic when sample sizes are low, as it can introduce noise and degrade the performance of the global model. Therefore, we introduce an exponential-based aggregation method to calculate the weight \(\beta_n\):

\begin{equation}
    \beta_n = e^{\lambda \times \frac{sample_n}{\sum_{n=1}^N sample_n}}
\end{equation}

where \({sample}_n\)  represents the number of samples for the \(n\)th STA, \(N\) is the total number of STAs participating in the aggregation, and \(\lambda\) is an adjustable parameter. The calculation of the aggregated model weight \(W\) can then be expressed as:

\begin{equation}
    W = \sum_{n=1}^N \beta_n \omega_n
\end{equation}

This exponential aggregation method can effectively reduce the negative impact of STAs with few samples on the aggregated model, thereby improving the stability and accuracy of the aggregated model.


\section{Simulation Result}
We constructed an 802.11ax Wi-Fi network topology environment using Network Simulator (ns-3) and implemented the E-FRL algorithm with PyTorch, utilizing ns3-ai \cite{b21} as the interface between the algorithm and the Wi-Fi network environment. The main simulation parameters in ns-3 are shown in Table \ref{table:ns3_parameters}, and the hyperparameters of the DDPG algorithm are listed in Table \ref{table:efrl_hyperparameters}.

We validated the algorithm’s performance by evaluating the system’s total throughput and average MAC delay. Two scenarios were considered: a static scenario where the distance between the AP and STA varies and the upstream traffic arrival rate of the STA is fixed; and a dynamic scenario where STAs with fixed and random traffic coexist. We first compared the proposed algorithm in a static scenario with a baseline without model pruning strategy to demonstrate the impact of the model pruning strategy on the FL training process and model performance, laying the foundation for subsequent dynamic scenario testing. Then, in the dynamic scenario, we compared the performance of the proposed E-FRL scheme with three baselines: Average Federated Reinforcement Learning (A-FRL), DRL, and RTS/CTS. A-FRL uses the FedAvg algorithm for FL model aggregation. In the DRL algorithm, each STA performs distributed training completely independently, without any FL aggregation process. RTS/CTS uses the traditional DCF method, sending RTS and CTS frames to reduce collisions and improve data transmission reliability.

\begin{table}[!t]
\caption{ns-3 Simulation Parameters}
\centering
\label{table:ns3_parameters}
\begin{tabular}{|c||c|}
\hline
\textbf{Parameter} & \textbf{Value} \\ \hline
MCS index & HE MCS(0-11) \\ \hline
Frequency & 5GHz \\ \hline
Bandwidth & 80MHz \\ \hline
Payload size & 1472 bytes \\ \hline
Transmitted power & AP: 23 dBm, STA: 20 dBm \\ \hline
Path loss model & LogDistancePropagationLossModel \\ \hline
Distance between STA and AP & [0.5, 30]m \\ \hline
Traffic type & UDP \\ \hline
\end{tabular}
\end{table}

\begin{table}[!t]
\caption{Hyper-Parameters of E-FRL Scheme}
\centering
\label{table:efrl_hyperparameters}
\begin{tabular}{|c||c|}
\hline
\textbf{Parameter} & \textbf{Value} \\ \hline
Soft update factor  & 0.001 \\ \hline
Learning rate & 0.002 \\ \hline
Batch size & 64 \\ \hline
Experience buffer & 2000 \\ \hline
Time step & 20ms \\ \hline
FL interaction period & 2.5s \\ \hline
Simulation time & 20s \\ \hline
\end{tabular}
\end{table}
\subsection{Static Scenario}
\begin{figure}[!t]
\centering
\includegraphics[width=2.3in]{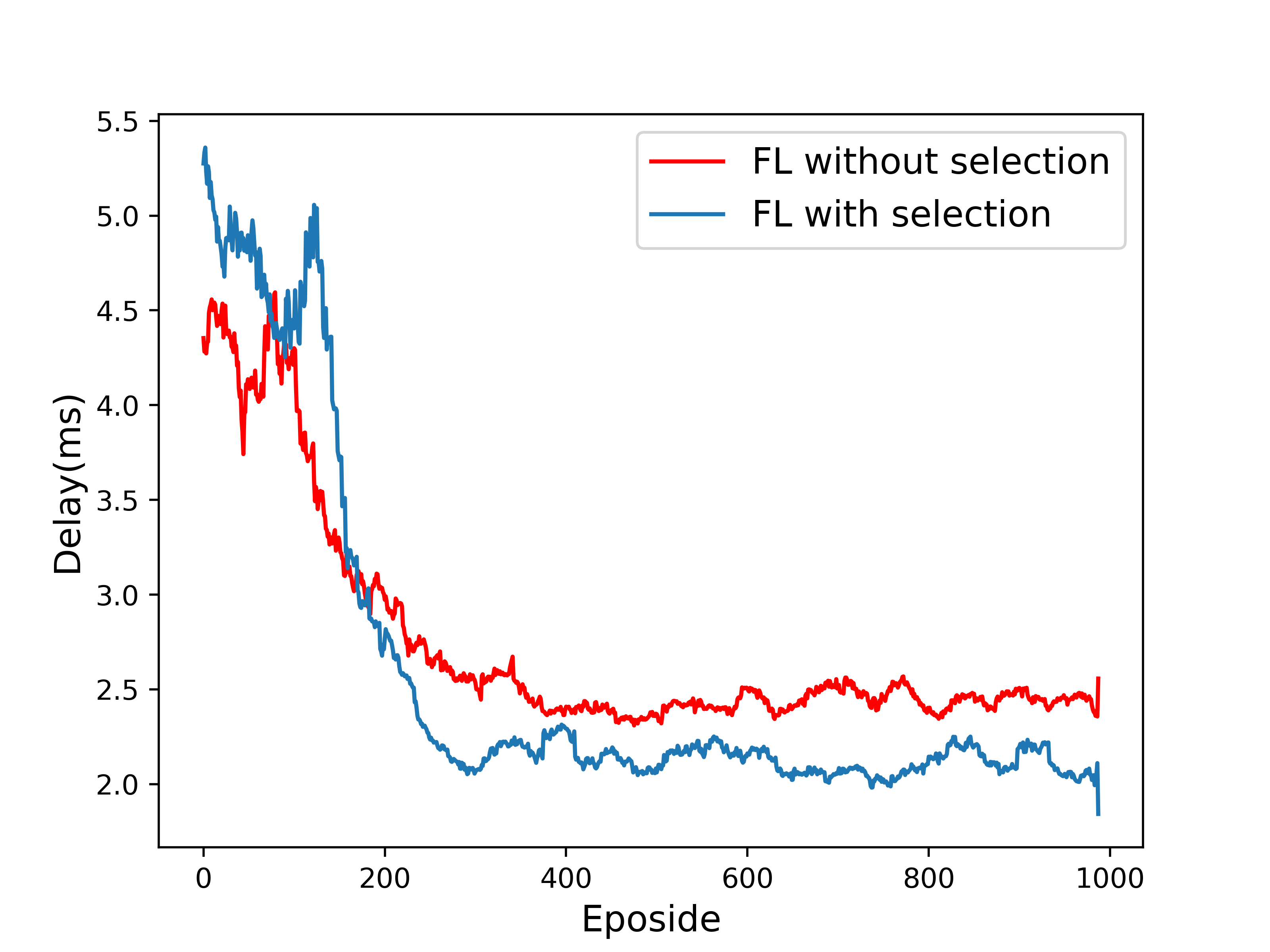}
\caption{Static scenario delay performance vs. Training eposides.}
\label{fig:static_1}
\end{figure}

\begin{figure}[!t]
\centering
\includegraphics[width=0.3\textwidth,height=0.167\textheight]{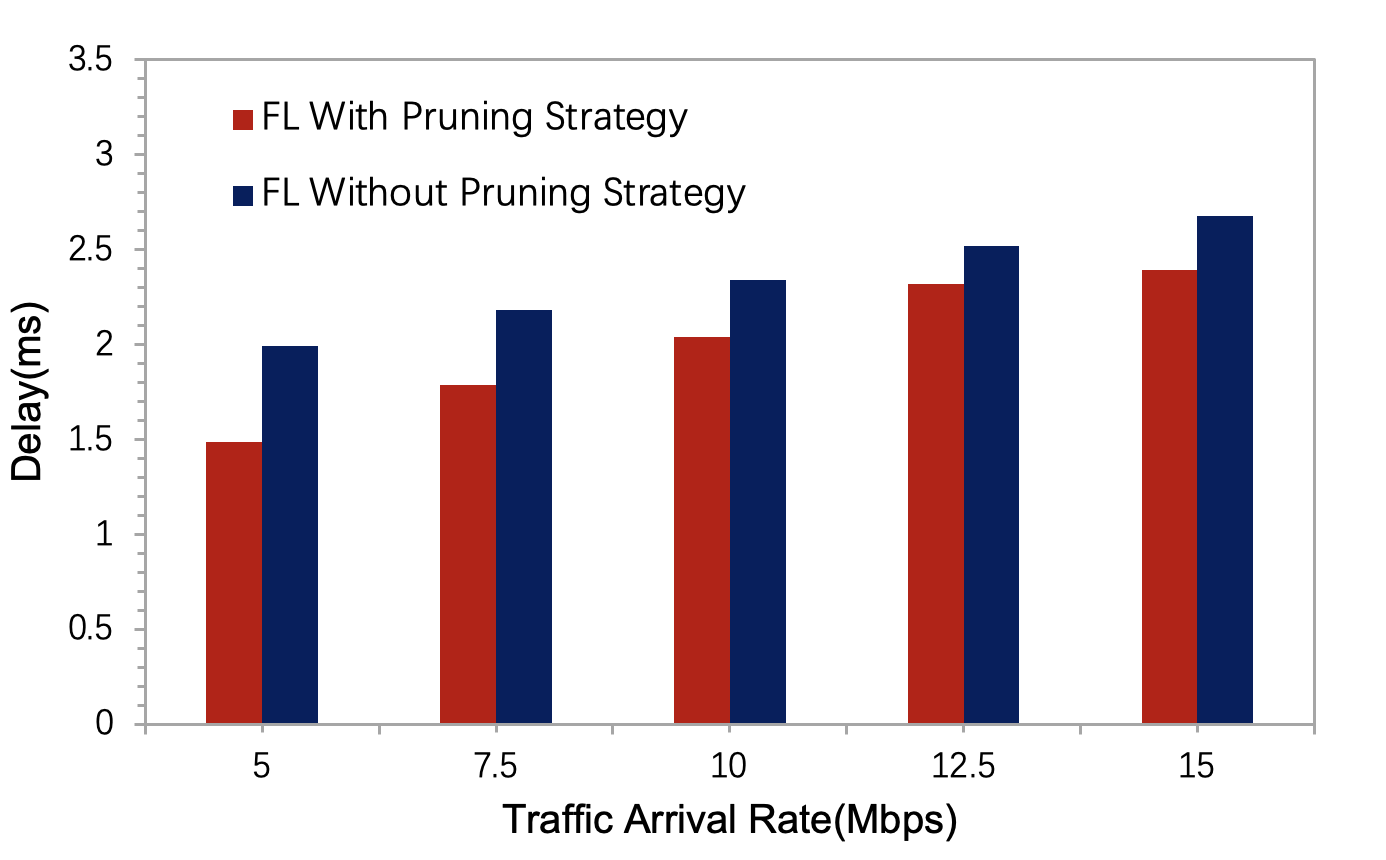}
\caption{Delay comparison at different traffic arrival rates in static scenario.}
\label{fig:static_2}
\end{figure}

Fig.~\ref{fig:static_1} shows the comparison of the proposed algorithm and baseline algorithms in terms of global MAC average delay when the traffic arrival rate is 10 Mbps, and the number of STAs is 10. As training time increases, the delay gradually decreases and stabilizes. The proposed algorithm, after introducing the model puncturing mechanism, shows a significant reduction in MAC average delay compared to the baseline. Fig.~\ref{fig:static_2} shows the comparison of global MAC average delay performance between the two schemes at different traffic arrival rates. As the traffic arrival rate increases, the MAC average delay also increases. However, the proposed scheme, by selecting STAs that are closer and have better channel conditions to participate in model aggregation, improves the performance of the global model. The MAC average delay in the proposed scheme is reduced by 7.96\% to 25.24\% compared to the baseline algorithms.


\subsection{ Dynamic Scenario}

\begin{figure}[!t]
\centering
\includegraphics[width=2.22in]{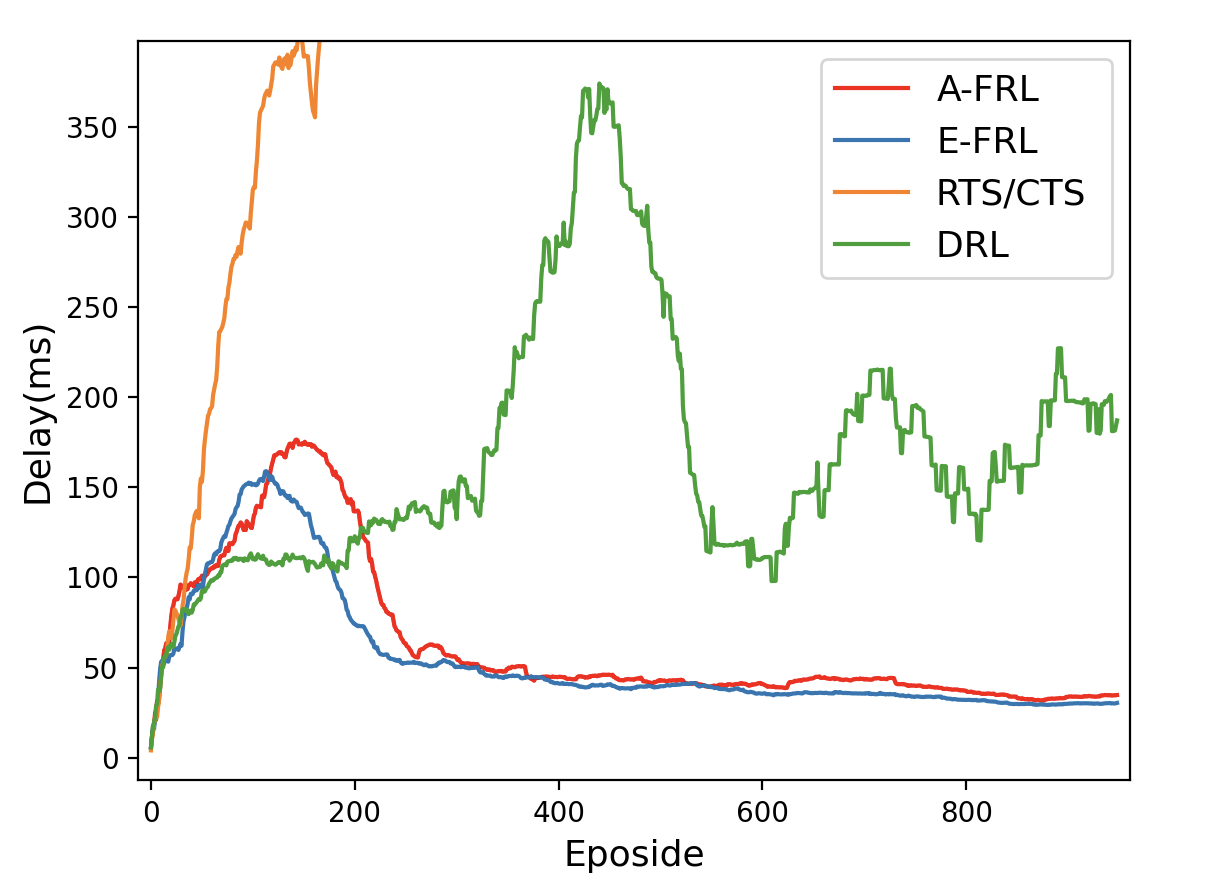}
\caption{Dynamic scenario delay performance vs. Training eposides.}
\label{fig:dynamic_1}
\end{figure}

\begin{figure}[!t]
\centering
\includegraphics[width=2.35in]{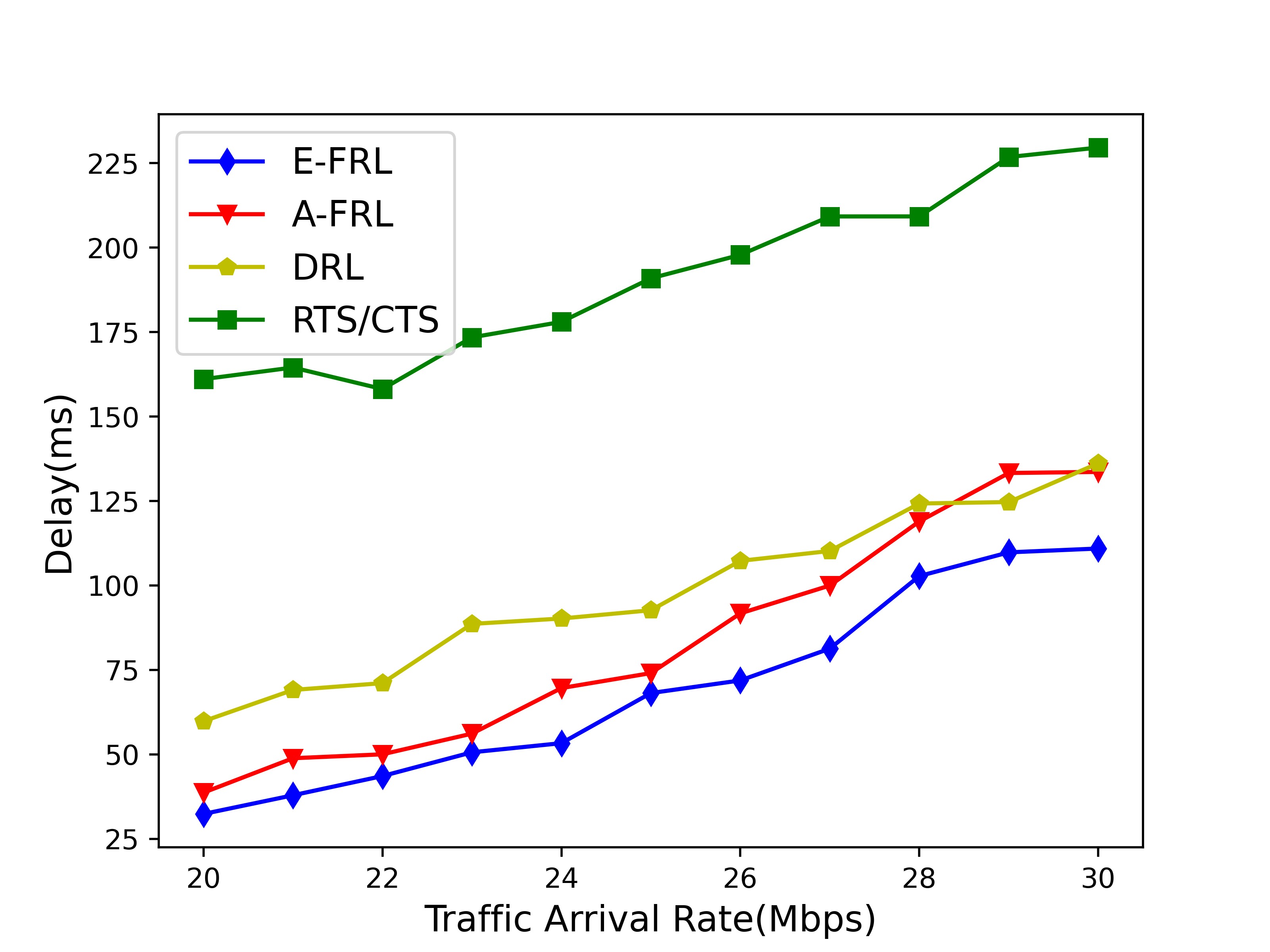}
\caption{Delay comparison at different traffic arrival rates in dynamic scenario.}
\label{fig:dynamic_2}
\end{figure}

\begin{figure}[!t]
\centering
\includegraphics[width=2.35in]{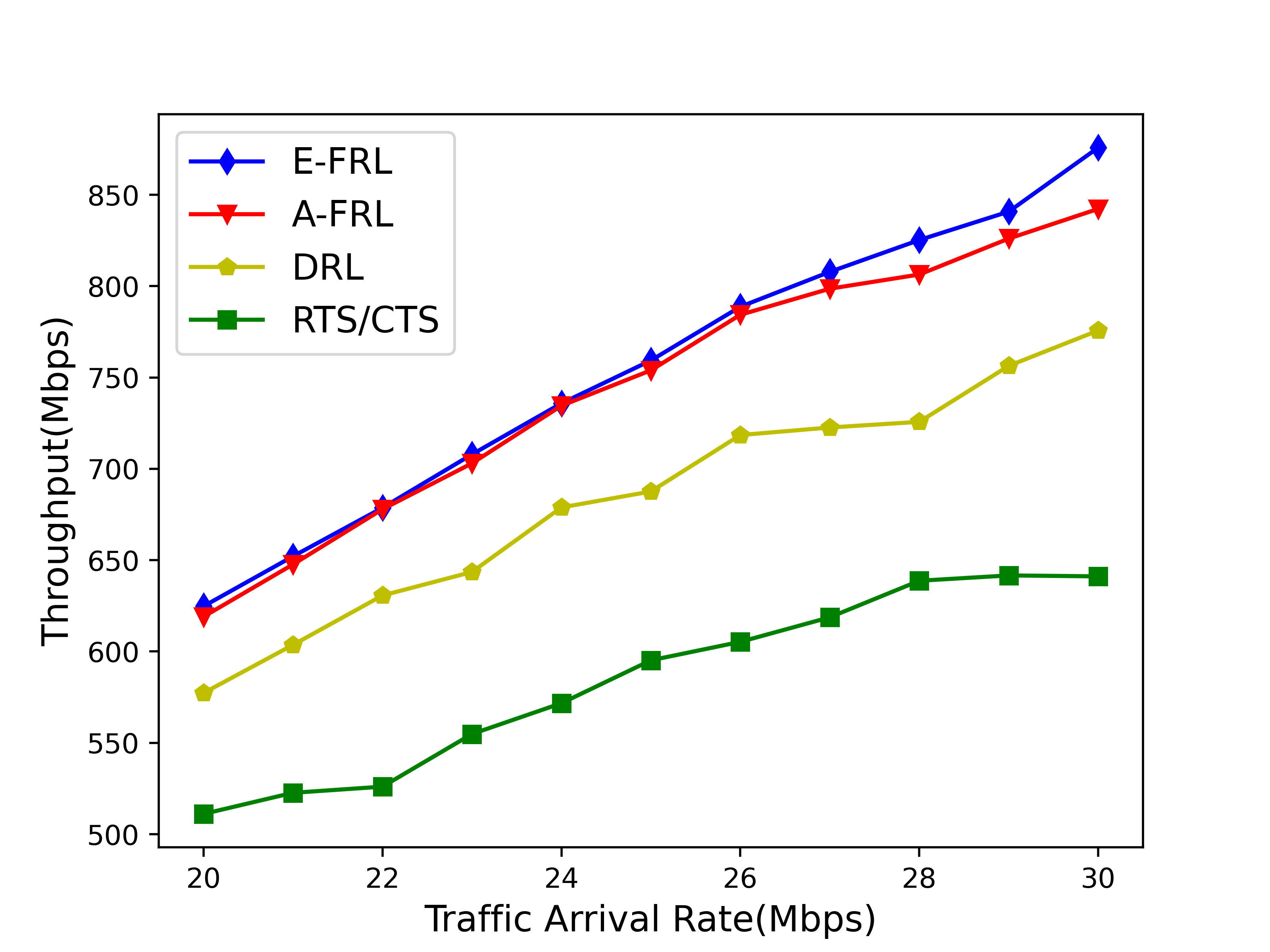}
\caption{Throughput comparison at different traffic arrival rates in dynamic scenario.}
\label{fig:dynamic_3}
\end{figure}

Assume that 60 STAs are associated with the same AP. Among them, 15 STAs have a fixed traffic arrival rate, while the remaining 45 STAs have a random traffic arrival rate. Over a certain period, the number of STAs with data transmission demands remains stable, keeping the level of competition in the network steady. Fig.~\ref{fig:dynamic_1} shows the delay performance variation with training rounds in the dynamic scenario. In DRL, due to the lack of FL, STAs rely only on local estimation for model training, leading to large fluctuations in global MAC average delay. In contrast, algorithms incorporating FL gradually stabilize in the later stages of training, and the proposed algorithm outperforms the A-FRL algorithm in MAC average delay due to a more reasonable aggregation algorithm. Fig.~\ref{fig:dynamic_2} and Fig.~\ref{fig:dynamic_3} show the MAC average delay and global throughput performance comparison under different traffic arrival rates in the dynamic scenario. As the traffic arrival rate increases, both the MAC average delay and throughput of the proposed and baseline algorithms increase significantly. However, the proposed scheme, benefiting from the application of the FL framework and the adoption of a reasonable pruning strategy and aggregation weight calculation method, significantly outperforms several baseline schemes. Compared to A-FRL, MAC average delay is reduced by 12.87\% to 25.72\%, and throughput is increased by 3.98\%. Compared to DRL, the MAC average delay performance gain is up to 45.9\%, and the throughput gain is 14.2\%. Compared to RTS/CTS, delay is reduced by 51.59\% to 79.91\%, and throughput is increased by 36.5\%.



\section{Conclusion}
This paper proposes an intelligent channel competition access mechanism based on federated deep reinforcement learning, aiming to address the performance issues caused by channel competition in densely deployed WLANs. For a distributed Wi-Fi scenario with a single AP and multiple STAs, by introducing FL to enhance overall system fairness, the proposed method also fully considers dynamically changing channel quality and STA uplink traffic, introducing a pruning strategy and an exponential aggregation algorithm to effectively reduce unnecessary global model training, lower communication overhead, and improve the quality of data during training. This ensures the stability and accuracy of model training. Simulation results show that compared with traditional machine learning methods, the proposed scheme not only reduces MAC average delay but also demonstrates higher adaptability and robustness in practical applications.

\vfill

\end{document}